\documentclass[12pt]{iopart}

\usepackage{graphicx}
\usepackage{iopams}
\begin{document}
\title{Mapping the lattice-vibration potential using terahertz pulses}

\author{C L Korpa$^1$, Gy T\'oth$^2$ and J Hebling$^{1,2}$}

\address{$^1$ Institute of Physics, University of P\'ecs, Ifj\'us\'ag \'utja 6, 7624 P\'ecs, Hungary}
\address{$^2$ MTA-PTE High-Field Terahertz Research Group, Ifj\'us\'ag \'utja 6, 7624 P\'ecs, Hungary}
\eads{korpa@fizika.ttk.pte.hu, tothgy@fizika.ttk.pte.hu, hebling@fizika.ttk.pte.hu}

\begin{abstract}
We develop a method for mapping the anharmonic lattice potential using the time-dependent
electric field of the transmitted pulse through thin sample supported by a substrate of non-negligible  thickness. Assuming linear propagation in the substrate we fully take into account internal reflection in it while the sample can show arbitrary nonlinear response. We examine the effect of frequency averaging appropriate for broad-band pulse and compare the results taking into account the full frequency dependence.
We illustrate the
procedure applying it to a model based on recently observed ferroelectric soft-mode nonlinearity in SrTiO$_3$.
\end{abstract}

%\pacs{42.25.Bs, 42.65.An, 63.20.Ry}

\vspace{2pc}
\noindent{\it Keywords}: anharmonic lattice vibration, terahertz radiation

\submitto{\JPB}
\maketitle

\section{Introduction}
Development and application of sources of short, intense radiation in the terahertz region \cite{Hebling-2009a,Katayama-2008,Tanaka-2012,Kampfrath-2013,Terahertz-2013} makes possible observation of vibrational and electronic excitations in various materials \cite{Ryder-2014,Mankowski-2014,Kreilkamp-2016,Hermann-2016,Bowlan-2017,Poperezhai-2017}
and quantification of nonlinear effects in pulse propagation \cite{Gaal-2008,Korpa-2016,Nguyen-2007}. This
offers the possibility of studying material properties responsible for nonlinear effects in the frequency
region previously not accessible and potentially even manipulating these properties by induced phase transition.
Driving large-amplitude vibrations by resonant low-frequency electromagnetic radiation allows efficient studying
of anharmonic lattice oscillations and thus direct determination of the corresponding lattice potential \cite{Tanaka-2012}. If the effect of damping force on  oscillations is substantial, as expected in case of large amplitudes, the term modelling it has to be introduced explicitly since it cannot be attributed to a potential term.

In this paper we develop a systematic procedure for mapping the vibrational lattice potential by comparing the time-dependent shape of few-cycle terahertz pulse transmitted through a thin sample and supporting substrate with the shape of the reference pulse transmitted through the substrate only.
In section 2 we present a theoretical analysis of pulse propagation using the transfer-matrix method adapted to studying thin polarization sheets \cite{Felderhof-1987}. Taking into account nonlinear effects in the thin sample is straightforward since the negligible thickness eliminates the need for considering nonlinear propagation. In section 3 we provide a detailed numerical analysis of single-cycle terahertz pulses traversing a thin SrTiO$_3$ sample exciting its anharmonic lattice oscillation. SrTiO$_3$ was chosen since the shape of input single-cycle terahertz pulse and the transmitted pulse has recently been accurately measured \cite{Tanaka-2012}, and since SrTiO$_3$ is an important material in microwave technique. In both sections we perform the analyses based on the full frequency spectrum of pulses and also using frequency-averaged expressions for pulse propagation. The latter is appropriate for a sufficiently broad frequency spectrum of the pulse in which case periodic functions of the phase shift of back-and-forth propagation in the substrate $2kL$, where $L$ is the substrate thickness and $k$ the wave number, go through at least a complete cycle in the pulse band \cite{GT-1957}. This simplifies the analysis considerably by eliminating the need for detailed knowledge of pulses' frequency spectra but the  accuracy hinges on sufficiently broad and smooth frequency spectra of pulses and parameters of system not changing significantly in the relevant frequency interval.

\section{Theoretical considerations}
\label{s1}
Our aim is to extract information about material polarization comparing the time dependence of
few- or single-cycle pulse transmitted through sample and substrate with that of transmitted through
substrate only. We assume thin sample with thickness much smaller than the shortest wavelength
present in the pulse since that leads to great simplification of the analysis made possible by considering  homogeneous fields in the sample. The thickness of the substrate to which the sample is attached we consider to be arbitrary, with limitation imposed only by a requirement that nonlinear effects in
the pulse propagation through the substrate remain negligible. The possibility of such an arrangement is supported by observing the practically indistinguishable shape of pulses traversing only the substrate, but having amplitudes differing by a factor of eight and in the case of more intense pulse causing anharmonic vibrations in the sample \cite{Tanaka-2012}. In this way an analytic treatment of the
pulse propagation can be achieved with arbitrary nonlinearity in the sample.
Following the substrate we assume a pulse propagating only in the direction of the incident wave while
internal reflection in the substrate is fully taken into account.

\begin{figure}[h]
\centering
\includegraphics[width=60mm]{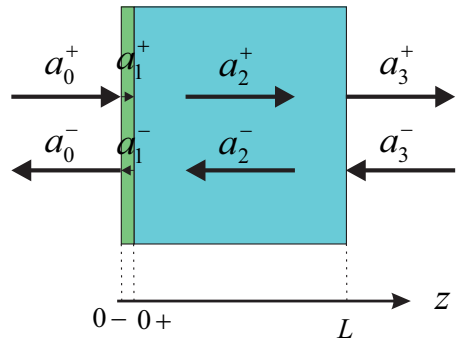}
\caption{Schematic representation of the electric-field components propagating in the sample (at $z=0$), substrate extending until $z=L$, and their  vicinity.
\label{fig0}}
\end{figure}

By avoiding nonlinear propagation we can use the transfer-matrix method adapted to consideration of thin
polarization sheets \cite{Felderhof-1987}.
We represent the physical electric field ${\bi E}_{\rm{ph}}({\bi r},t)$ by analytic complex field ${\bi E}({\bi r},t)$ whose Fourier transform contains only positive frequencies \cite{Haykin-2001,Conforti-2010}:
\begin{equation}
{\bi E}_{\rm{ph}}({\bi r},t)=\frac{1}{2}\left[ {\bi E}({\bi r},t)+\mbox{c.c.}\right]
\end{equation}
and similarly the magnetic field ${\bi B}_{\rm{ph}}({\bi r},t)$. We consider a linearly polarized pulse
propagating parallel and antiparallel to the z axis, with normal incidence on the sample and substrate.
The Fourier component $E(z)$ we write as
\begin{equation}\label{}
  E_j(z)=a_j^+\, \exp{(\rmi k_j z)}+a_j^-\, \exp{(-\rmi k_j z)},
\end{equation}
with $k_j=\omega n_j/c$ and $n_j$ being the refractive index of medium denoted by index j and
$a_0^+ (a_0^-)$ characterizing the incoming (reflected) component. The sample thickness $d$ satisfies the relationship $d\ll \lambda_{\rm{min}}$ where $\lambda_{\rm{min}}$ is the minimal wavelength present in the sample.
In the analysis the sample is assumed to be located between $z=0-$ and $z=0+$ planes, i.e.\ having negligible thickness, and is
attached to a substrate of thickness $L$ as schematically shown in Fig.~\ref{fig0}.
Using the continuity of the tangential component of the electric field and the jump condition for the magnetic field $B$ (assuming non-magnetic material):
\begin{equation}%\label{}
  \Delta B=B(0+)-B(0-)=\mu_0\, d\, J_p=\mu_0\, d\, \sigma(\omega,E)\, E,
\end{equation}
where $J_p$ is the polarization current density and the conductivity $\sigma$ depends also on the
electric field due to nonlinearity of the polarization sheet, we write:
\begin{eqnarray}\label{trans0}
% \nonumber % Remove numbering (before each equation)
  a_1^+ +a_1^- &=& a_0^++a_0^- \nonumber \\
  a_1^+ -a_1^- &=& a_0^+-a_0^- -c \Delta B.
\end{eqnarray}
We remark that the dependence of the conductivity on the electric field means that one should understand the following expressions containing the conductivity $\sigma$ (or $\beta$ as introduced in the following) as implicit relationships between quantities of the expression where the electric field in the sample is also present. This means that these expressions are not useful for direct calculations. However, as shown in the following, see Eq.\ (\ref{polexp}), it turns out that we need to use only an expression in which the conductivity is multiplied by the electric field in the sample giving the polarization current density which is, in time domain, the time derivative of the polarization. From time dependence of the polarization we can determine the charge displacement and the time-dependent lattice force.

Solving Eqs.\ (\ref{trans0}) for $a_1^+,a_1^-$ we obtain the transfer matrix for the polarization sheet
\cite{Felderhof-1987}:
\begin{equation}%\label{}
  S_0=\left( \begin{array}{cc}
               1+\beta & \beta  \\
               -\beta  & 1-\beta
             \end{array} \right),
\end{equation}
where $\beta=-d\,\sigma(\omega,a_0^++a_0^-)/(2c\epsilon_0)$ and we used that $E(0)=a_0^++a_0^-$.

The transfer matrix for the interface of polarization sheet and substrate with refraction index $n$ at $z=0+$ can be written as:
\begin{equation}%\label{}
  S_1(z)=\left(
\begin{array}{cc}
 e^{\rmi k z} & 0 \\
 0 & e^{-\rmi k z}
\end{array}
\right)
  \cdot
  \left(
\begin{array}{cc}
 \frac{n+1}{2 n} & \frac{n-1}{2 n} \\
 \frac{n-1}{2 n} & \frac{n+1}{2 n}
\end{array}
\right),
\end{equation}
with $k=\omega n/c$.
Finally, for the interface of the substrate at $z=L$ we have for the transfer matrix:
\begin{equation}%\label{}
  S_2=\left(
\begin{array}{cc}
 \frac{n+1}{2} & \frac{1-n}{2} \\
 \frac{1-n}{2} & \frac{n+1}{2}
\end{array}
\right).
\end{equation}
The electric field leaving the substrate at $z=L+0$ is then given by:
\begin{equation} %\label{}
  \left( \begin{array}{c}
           a_3^+ \\
           a_3^-
         \end{array} \right) =S_2\cdot S_1(L)\cdot S_0 \cdot
         \left( \begin{array}{c}
           a_0^+ \\
           a_0^-
         \end{array} \right).
\end{equation}
Imposing the condition $a_3^-=0$ determines $a_0^-$ in terms of $a_0^+$ and thus gives the electric field leaving the substrate in terms of the incoming one:
\begin{equation} \label{efull}
  a_3^+=\frac{4 n e^{i k L}}{(n+1) (n+1-2 \beta)-e^{2 i k L} (n-1) (n-1+2 \beta )}\,a_0^+,
\end{equation}
while the electric field in the sample is given by:
\begin{equation}\label{esample}
 a_1^++a_1^-= a_0^++a_0^-=\frac{a_0^+}{\frac{1}{2} (-2 \beta -n+1)+\frac{n (n+1)}{(n-1) \exp{(2 i k L)}+n+1}}.
\end{equation}

For pulses with wide bandwidth expressions (\ref{efull}) and (\ref{esample}) need averaging over the relevant spectral distributions.
For few-cycle pulses and substrate thickness $L$ comparable to the wavelength corresponding to the central frequency of the pulse one can expect to get reasonable agreement by doing analytic averaging over complete cycle
of $2kL$ with constant weight. Thus, we also present expressions corresponding to such an analytic averaging and call those results cycle averaged ones and denote them by overline. We remark that such cycle averaged results for quite a few, but not all expressions are identical to the ones obtained by neglecting the internal reflection in the substrate.
We obtain:
\begin{equation}\label{efullav}
  \overline{\frac{a_3^+}{a_0^+}}=\frac{4n}{(n+1)(n+1-2\beta)}
\end{equation}
and
\begin{equation}\label{esampleav}
  \overline{\frac{a_0^++a_0^-}{a_0^+}}=\frac{2}{n+1-2\beta},
\end{equation}
where we assume that the refractive index $n$ and parameter $\beta$ do not change significantly with the frequency variation corresponding to the period of $k L$.

By taking the absolute value squared of eq.\ (\ref{efull}) we obtain the intensity transmission $T$ and reproduce  expression (A-1) of Glover and Tinkham \cite{GT-1957} for the case of linear response of the sample (in their notation $-2\beta=g-\rmi b$). The ratio of the electric field traversing the sample and substrate and that going through the substrate only after cycle averaging is:
\begin{equation}\label{ratioav}
  \overline{\frac{a_3^+(\beta)}{a_3^+(\beta=0)}}=\frac{n+1}{n+1-2\beta}=\frac{1}{1-\frac{2\beta}{n+1}},
\end{equation}
which is the well-known Tinkham formula \cite{Tu-2003,Tinkham-1956} obtained by neglecting internal reflection in the substrate. However, we point out that neglecting internal reflection in substrate can lead to significantly different result from the cycle averaged one for the pulse passing through the substrate only:
\begin{equation}\label{ratioavsub}
  \overline{\frac{a_3^+(\beta=0)}{a_0^+(\beta=0)}}=\frac{4n}{(n+1)^2}=\frac{2}{n+1}\cdot \frac{2}{1+1/n},
\end{equation}
since the second factor in the last equality is significantly larger than one if the refraction index substantially exceeds one. Thus, when only the ratio of the transmitted pulses through sample + substrate and substrate only is used for analysis the Tinkham formula for a wide-band pulse is adequate. If the properties of the pulse transmitted through the substrate are used, for example in order to determine the incoming electric field, the full result with internal reflection should be used.

To determine the electric field in the film in terms of the pulse transmitted through sample and substrate we combine Eqs.\ (\ref{efull}) and (\ref{esample}) to obtain:
\begin{equation}\label{efilmgen}
  \frac{a_3^+}{a_0^++a_0^-}=\frac{2n\,\exp{(\rmi k L)}}{n+1+(n-1)\exp{(2\rmi k L)}}.
\end{equation}
Eq.\ (\ref{efilmgen}) is useful since it does not contain the nonlinear term $\beta$ and thus allows for explicit determination of the field in the sample. This result is in acccordance with the assumption of  very thin sample and propagation taking place only in the substrate. The cycle-averaged result is:
\begin{equation}\label{efilmav}
 \overline{ \frac{a_3^+}{a_0^++a_0^-}}=\frac{2n}{n+1},
\end{equation}
where we dropped the phase factor in the numerator leading to a simple time shift.

Now we can rewrite Eq.\ (\ref{esample}) such that $\beta$ multiplying $a_0^++a_0^-$ is separated, thus relating the electric field in the sample to the incoming field and the polarization-current density:
\begin{equation} \label{polexp}
  2a_0^+=-2\beta (a_0^++a_0^-)+\frac{(n+1)^2-(n-1)^2\,\exp{(2\rmi k L)}}{n+1+(n-1)\,\exp{(2\rmi k L)}}\,(a_0^++a_0^-).
\end{equation}
Transforming to the time domain we get the time derivative of the polarization:
\begin{equation} \label{polexptime}
  2E_{\rm{in}}(t)=\frac{d}{\epsilon_0 c}\frac{\partial P(t)}{\partial t}+E_f(t),
\end{equation}
where $E_f(t)$ is the real part of the inverse Fourier transform of the second term on the right-side of Eq.\ (\ref{polexp}). We remark that in the cycle-averaged approximation that term reduces to $(n+1)\,E_{\rm{film}}(t)$ and (\ref{polexptime}) takes the form which was used in Ref.~\cite{Tanaka-2012}. Since the polarization is proportional to the parameter $Q(t)$ characterizing charge displacement from (\ref{polexptime}) we can obtain the time derivative $\dot{Q}(t)$. In order to get information about the restitutive force $F_r(\dot Q,Q)$, for which we assume to contain a damping term depending on time derivative of displacement, we write the equation of motion:
\begin{equation} \label{eqmotion}
  M\,\ddot{Q}=e^\star \, E_{\rm{film}}(t)+F_r(\dot Q,Q),
\end{equation}
where $M$ is the relevant mass and $e^\star$ the effective charge. For the force $F_r(\dot Q,Q)$ it is reasonable to assume a form
\begin{equation} \label{frform}
  F_r(\dot Q,Q)=-\alpha\, \dot Q +f_r(Q),
\end{equation}
with $f_r(0)=0$ and $\alpha$ a positive constant or possibly also depending on $Q$ as taken in Ref.\ \cite{Tanaka-2012}.

In the equation of motion (\ref{eqmotion}) we can compute the left-hand side by differentiation of $\dot{Q}(t)$ obtained from the time derivative of polarization and using the calculated electric field in the sample based on Eq.\ (\ref{esample}) we determine the time dependence of the force $F_r$. In order to get the displacement dependence of the force $F_r$ we use the time dependence of $Q(t)$ (obtained by integrating $\dot Q(t)$) and then make a parametric plot $(Q(t),F_r(t))$. Subtracting the value of $F_r$ at $Q=0$ we get the function $f_r(Q)$ providing the lattice-vibration potential $V(Q)$ by integration. The subtraction constant can be fit to $-\alpha\, \dot Q$ by using $\dot Q(t_0)$ where $Q(t_0)=0$. Alternatively, as shown in the next section, one may use a multipoint fit to $F_r(\dot Q,Q)$ exploiting the fact that $f_r(Q)$ is linear for small values of displacement $Q$.

In the next section we illustrate the above presented method by a detailed numerical analysis of a model based on the recent  measurement reported in Ref.~\cite{Tanaka-2012}.

\section{Numerical modeling}
\label{numerics}
In this section we perform an analysis of lattice oscillations in SrTiO$_3$ whose ferroelectric soft mode was impulsively driven to anharmonic regime using terahertz pulses in Ref.\ \cite{Tanaka-2012}. SrTiO$_3$ has a highly anharmonic mode with very large oscillator strength at low temperature in the terahertz region \cite{Barker-1966,Vogt-1995} and low potential barrier to ferroelectric phase transition which makes it especially convenient for investigation of possible terahertz induced phase transformations.

For our analysis we use
a model in which the reference pulse is taken from data presented in Ref.\ \cite{Tanaka-2012}.
This provides a pulse with realistic frequency spectrum suitable for modeling experimental circumstances.
In that experiment transmission of single-cycle terahertz pulse through a 300nm thick SrTiO$_3$ sample and MgO substrate with 0.5mm thickness was used. We take the measured shape of the reference pulse passing through the substrate only and use it to calculate the shape of the pulse passing through both sample and substrate applying the lattice oscillation model obtained in \cite{Tanaka-2012}.

\begin{figure}[h]
\centering
\includegraphics[width=80mm]{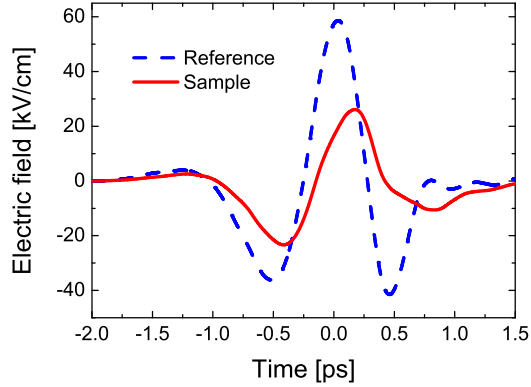}
\caption{Time dependence of the electric field of the reference pulse passing through substrate (dash line) and calculated pulse transmitted through both sample and substrate (solid line). The normalization is obtained using the maximum incoming field value at the sample of 80 kV/cm \cite{Tanaka-2012}.
\label{fig1}}
\end{figure}

By using a simulated input for the sample response based on a definite lattice-oscillation model we can analyze the accuracy of the developed method in reproducing the lattice potential from the response of the sample and the observed reference pulse.
In Fig.~\ref{fig1} we show the reference and the sample fields, the latter resulting from a calculation using the lattice vibration parameters obtained in Ref.~\cite{Tanaka-2012}. We use the cycle averaged expressions when calculating the shape of the pulse transmitted through both sample and substrate exploiting the reference-pulse shape and a knowledge of the lattice potential since using expressions with the full frequency dependence of the pulses does not allow such a direct calculation.
However, working in the opposite direction, i.e.\ calculating the lattice force from the reference and sample fields is possible for both approaches as outlined above. Thus, for consistent use of our model input $E_{\rm{sample}}(t)$ we need to use in the analysis the cycle averaged expressions. We also perform the analysis with full frequency dependence in order to get a (semi) quantitative estimate of the effect of cycle averaging for the studied case.
The normalization is based on the  maximum incoming field value at the sample of 80 kV/cm as stated in \cite{Tanaka-2012} and using the cycle-averaged expression (\ref{ratioavsub}). For the refraction index of bulk MgO we used the constant value $n=3.1$ as suggested by the results of Ref.~\cite{Han-2008}.

\begin{figure}[h]
\centering
\includegraphics[width=80mm]{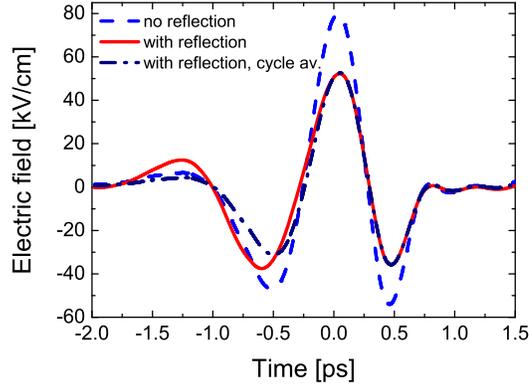}
\caption{The incoming electric field obtained from the reference pulse: solid line shows the full result, dash line corresponds to neglecting reflection in substrate, dash-dot line is for cycle averaged expression.
\label{fig2}}
\end{figure}

Using the reference pulse we can calculate the incoming electric field by expressing $a_0^+$ from Eq.~(\ref{efull}) with $\beta=0$. In Fig.~\ref{fig2} we show the result (solid line) together with the result of using the cycle averaged expression (\ref{ratioav}) with $\beta=0$ and also the curve obtained by neglecting the reflection in substrate. We observe considerable difference of the last curve compared to the other two which is a consequence of rather large substrate refraction index.

The next step is to use the relationship (\ref{polexptime}) to determine the polarization term. In the first analysis we use the cycle-averaged expressions leading to:
\begin{equation} %\label{}
  E_f(t)=(n+1)\,E_{\rm{film}}(t)=\frac{(n+1)^2}{2n}\,E_{\rm{sample}}(t).
\end{equation}
We also analyze the effect of using cycle averaging appropriate for broadband pulse in the case of substrate thickness at least comparable to wavelength of central frequency compared to the expression containing the full frequency dependence. Since using in the latter case the same $E_{\rm{sample}}(t)$ based on cycle averaged expressions gives not acceptable result for the displacement (not approaching zero after passage of the pulse) we adjust $E_{\rm{sample}}(t)$ in such a way that it leads to the same time-dependence of the field in the sample ($E_{\rm{film}}(t)$) as when using cycle averaging. This (obviously) cures the mentioned problem and the differences still present in shapes of other quantities can serve for estimating the importance of doing a full calculation compared to a simpler cycle averaged one in the analyzed case.

\begin{figure}[h]
\centering
\includegraphics[width=80mm]{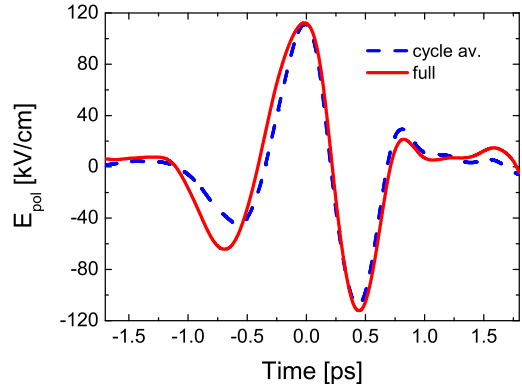}
\caption{The polarization contribution $\frac{d}{\epsilon_0 c}\frac{\partial P(t)}{\partial t}$ obtained using cycle averaged expressions (dash line) and full frequency dependence (solid line).
\label{fig3}}
\end{figure}

\begin{figure}[h]
\centering
\includegraphics[width=80mm]{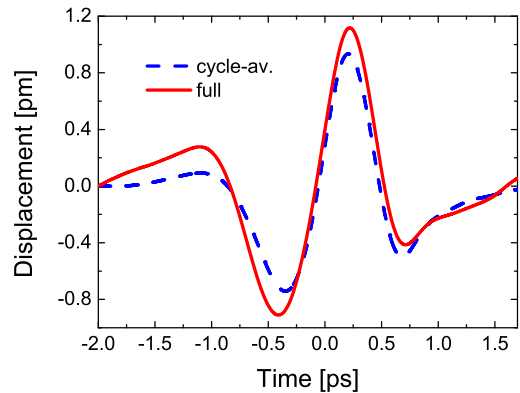}
\caption{The displacement parameter $Q(t)$. Dash line shows the result obtained with cycle averaging, solid line the full calculation.
\label{fig4}}
\end{figure}

\begin{figure}[h]
\centering
\includegraphics[width=80mm]{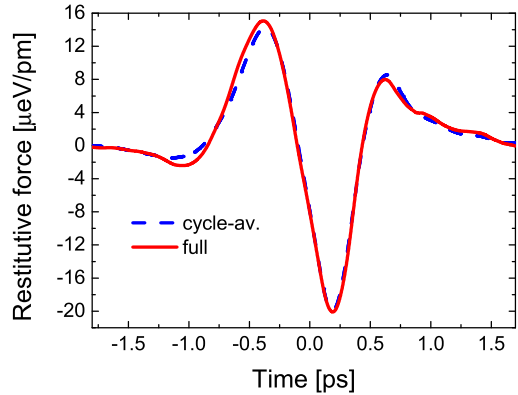}
\caption{Time dependence of the restitutive force $F_r(t)$. Dash line shows the result obtained with cycle averaging, solid line the full calculation.
\label{fig5}}
\end{figure}

\begin{figure}[h]
\centering
\includegraphics[width=80mm]{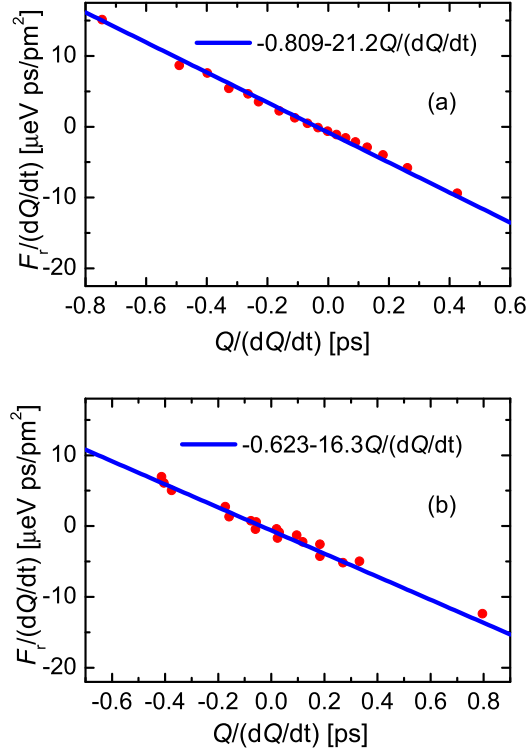}
\caption{Least-squares fits to $F_r(\dot Q,Q)/\dot Q$ for the case of using cycle averaged expressions (a), and full frequency dependence (b).
\label{figfit1}}
\end{figure}

\begin{figure}[h]
\centering
\includegraphics[width=80mm]{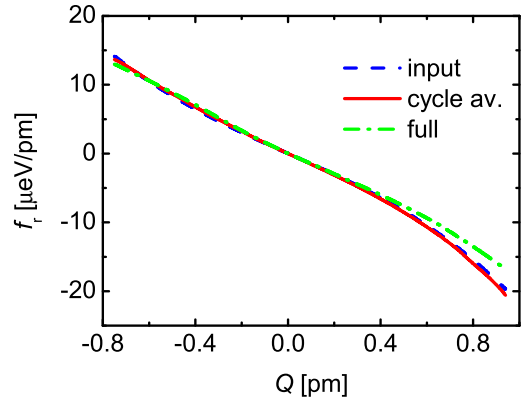}
\caption{The force $f_r(Q)$ for the case of using cycle averaged expressions (solid line) and full frequency dependence (dash-dot line) compared to the input (dash line).
\label{figforce}}
\end{figure}

\begin{figure}[h]
\centering
\includegraphics[width=80mm]{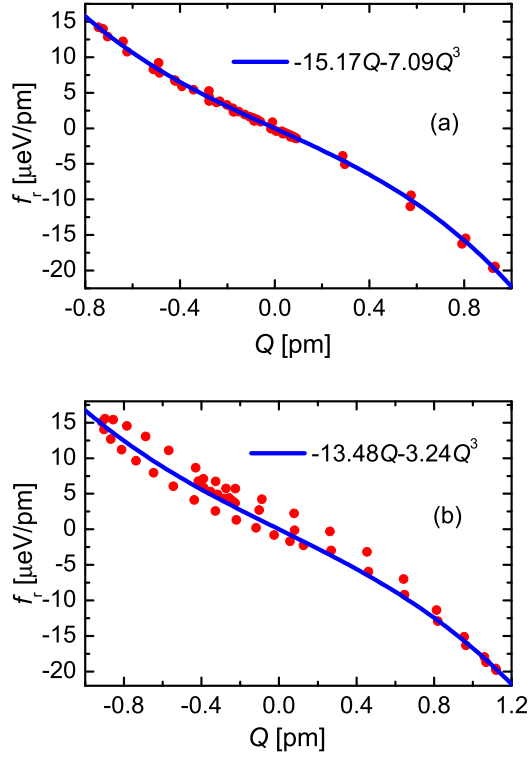}
\caption{Least-squares fits determining the force $f_r(Q)$ for the case of using cycle averaged expressions (a) and full frequency dependence (b).
\label{figfit2}}
\end{figure}

The result for the polarization contribution
\[
E_{\rm{pol}}=\frac{d}{\epsilon_0 c}\frac{\partial P(t)}{\partial t}
\]
is shown in Fig.~\ref{fig3}.
Using the data characterizing the crystal structure of SrTiO$_3$ given in \cite{Tanaka-2012} we can relate the polarization to the (effective) charge displacement $Q(t)$ whose time derivative is then related to polarization contribution: $E_{\rm{pol}}=22.36\,\dot Q(t)$, where $\dot Q(t)$ is in pm/ps and $E_{\rm{pol}}$ in kV/cm. Numerical integration gives the displacement $Q(t)$ as given in Fig.~\ref{fig4}.
From the equation of motion (\ref{eqmotion}) we can now get the time dependence of the restitutive force $F_r(\dot Q(t),Q(t))$ which can then be used to establish the dependence on $\dot Q$ and $Q$. Fiq.~\ref{fig5} shows the result for $F_r(t)$ after performing a numerical differentiation of $\dot Q(t)$.
To proceed for $F_r(\dot Q,Q)$ we use the form (\ref{frform}) which was assumed when determining the transmitted pulse $E_{\rm{sample}}(t)$ and now we want to extract the parameters determining the right-side of (\ref{frform}). We notice that it is delicate to get the $\dot Q$ coefficient $\alpha$ since that term in the equation of motion is overwhelmed by the polarization term $E_{\rm{pol}}$ which is also proportional to $\dot Q$. It turns out that instead of subtracting the force value at $Q=0$ it is more reliable to use a multiple-point least-squares fit for small $Q$ values such that we can neglect higher-order terms in $f_r(Q)$. It is convenient to do a first-order fit in $Q(t)/\dot Q(t)$ by writing:
\begin{equation} %\label{}
  F_r(t)/\dot Q(t)=-\alpha+\gamma \, Q(t)/\dot Q(t).
\end{equation}
Using values around $t=0$ where $\dot Q$ does not change sign we see from Fig.~\ref{figfit1} that for cycle averaged approach (a) the linear dependence is very satisfactory and for the full frequency dependence case (b) reasonable. We should not forget that for the latter case our input $E_{\rm{sample}}$ is not quite appropriate. Subtracting from $F_r(t)$ the term $-\alpha \dot Q(t)$ we get the time dependence of  $f_r(t)$ and using the calculated $Q(t)$ values we can simply deduce the functional dependence $f_r(Q)$.
The obtained results for the force $f_r(Q)$ are shown in Fig.~\ref{figforce}. We observe very good agreement with the input for cycle averaged result and noticeable deviation for the full calculation. One should not forget that in the latter case we are using an approximate input since it is not possible to directly calculate $E_{\rm{sample}}(t)$ for a given force and we use the deviations as a semi quantitative measure of the effect of cycle averaging.
For numerical comparison with the input expression for the force we proceed by doing a fit for $F_r-(-\alpha \dot Q)$ of the form $a\,Q+b\,Q^3$ and compare the parameters $\alpha, a,b$ with the input  values. In Fig.~\ref{figfit2} we show the used ``data'' points and the obtained parameter values, the upper panel referring to cycle averaged case, the lower panel to the use of full expressions.
We remark that in this case we used the time span of the whole pulse which means that same displacement regions are traversed more than once while for the plot of Fig.~\ref{figforce} we used only the largest monotonic piece corresponding to time interval from -0.3ps until +0.2ps.
The input
values corresponding to the parameters of Ref.\ \cite{Tanaka-2012} are: $\alpha=0.805\mu$eV ps/pm$^2$,
$a=-15.0\mu$eV/pm$^2$, $b=-6.79\mu$eV/pm$^4$, where we used the corresponding relationships: $\alpha=M\,\Gamma_0, a=-M\Omega^2, b=-M \lambda$ to the parameters $\Gamma_0, \Omega, \lambda$ of \cite{Tanaka-2012} and we
dropped the $Q$ dependence of $\alpha$ playing a minor role.
Comparing the values for parameter $\alpha$ obtained by the fit shown in Fig.~\ref{figfit1}a, 0.809$\mu$eV ps/pm$^2$, with the input value of 0.805$\mu$eV ps/pm$^2$ and the values for $a$ and $b$ given by the fit of Fig.~\ref{figfit2}a, -15.17$\mu$eV/pm$^2$ and
-7.09$\mu$eV/pm$^4$, with the input values (-15.0$\mu$eV/pm$^2$, -6.79$\mu$eV/pm$^4$)
we observe that for the cycle averaged case we reproduce the parameters $\alpha$ and $a$ with 1\% accuracy while $b$ is determined to around 5\%.
\begin{figure}[h]
\centering
\includegraphics[width=80mm]{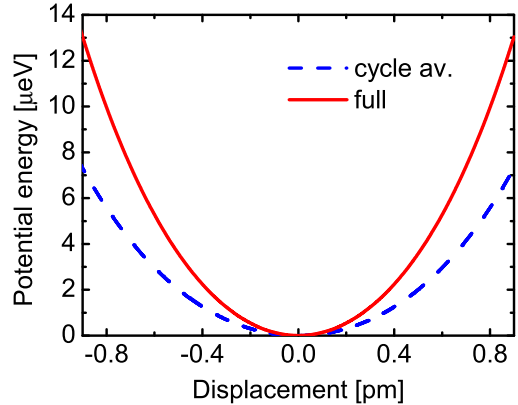}
\caption{Potential energy obtained by integrating the force $f_r(Q)$  shown in Fig.~\ref{figfit2} for the case of using cycle averaged expression (dash line) and full frequency dependence (solid line).
\label{figpot}}
\end{figure}
In Fig.~\ref{figpot} we show the potential energy resulting from the displacement dependence obtained by fits shown in Fig.~\ref{figfit2}. The potential used as input is on this figure indistinguishable from the cycle averaged result.

The results for the full approach indicate that the effect of replacing the complete frequency dependence of the expressions with the cycle averaged ones introduces significant differences in the parameters in the examined case. This is a consequence of pronounced difference in the shape of displacement's time dependence as shown in Fig.~\ref{fig4} although the time dependence of the force, Fig.~\ref{fig5}, is quite similar in both cases.  We remark that for larger substrate thickness this difference diminishes and an increase of substrate's thickness by a factor of four brings the cycle-averaged results very close to the full-frequency ones.

Depending on the nature of the investigated material and the amplitude of oscillations one can modify the functional dependence of the restitutive force $F_r(\dot Q,Q)$ on displacement $Q$ and its time derivative $\dot Q$ taking into account the relevant symmetry and possible higher order terms corresponding to larger out-of-equilibrium excitation without basically affecting the proposed method of analysis. However, the increased number of parameters to be determined would probably require parallel analysis of transmitted pulses having  different intensities in order to obtain well determined and reliable results.

\section{Conclusions}
\label{conc}
We studied the use of few-cycle terahertz pulses for calculating parameters of anharmonic lattice vibrations by comparing time dependence of the electric field of pulse transmitted through a thin sample and supporting substrate of arbitrary thickness with the time dependence of pulse transmitted through the substrate.
In order to avoid complicated analysis of nonlinear propagation in the sample we assume that its thickness is much smaller than the shortest wavelength present in the pulse thus assuring effectively uniform field in the sample.

For the propagation through substrate we assume that nonlinear effects can be neglected while in the sample they are fully taken into account. We also take into account the internal reflection in the substrate and analyze the effect of averaging over full cycle which is appropriate for pulses with broad-band spectrum and substrate whose thickness is at least comparable to the wavelength corresponding to the central frequency. We numerically analyzed a model based on recent measurement \cite{Tanaka-2012} and showed that the used input lattice force can be deduced with very good accuracy. We observed that internal reflection plays non-negligible role especially if the refractive index of substrate substantially exceeds one which is frequently the case for used materials in the terahertz region.
We expect that the method of analysis outlined above can be extended with suitable modifications to studies of the motion of bound electrons, assuming that much faster time variation of the electric field becomes amenable to experimental observation. Also, generalization to studying potential surfaces in three-dimensional space, while involving considerable complications due to the vector nature of displacement, polarization and electric field, seems accessible, possibly with judicious use of varying crystal orientation.

\ack

The present work was supported by the Orsz\'agos Tudom\'anyos Kutat\'asi Alapprogramok (OTKA, Hungary)
grant K109462 and is
dedicated to the 650th anniversary of the foundation of University of P\'ecs, Hungary.

\section*{References}

\bibliographystyle{iopart-num}
\bibliography{paper}

\end{document}